\documentclass[10pt]{llncs}
\usepackage[dvips]{epsfig}
\usepackage{a4}
\usepackage{cprog}
\usepackage{amssymb}
\def\FT{Fault-\kern-2pt Tolerant}
\def\WWW{World-\kern-1pt Wide Web}

\def\nmr{\hbox{$N$\kern-1pt MR}}

\font\sevenrm=cmr7
\font\nineit=cmti9
\let\mc=\ninerm % medium caps
\newbox\PPbox % symbol for ++
\setbox\PPbox=\hbox{\kern.5pt\raise1pt\hbox{\sevenrm+\kern-1pt+}\kern.5pt}
\def\PP{\copy\PPbox}
\def\CPLUSPLUS{{\mc C\kern-1pt\PP\spacefactor1000}}
\def\ICPLUSPLUS{{\nineit C\kern-1pt\mc\PP\spacefactor1000}}
\begin{document}
%
% Preamble:                     title, 
\title{The EFTOS Voting Farm:\\
A Software Tool for Fault Masking\\
in Message Passing Parallel Environments}
%
% Preamble:                             author list,
\author{Vincenzo De Florio, Geert Deconinck, Rudy Lauwereins}
%
% Preamble:                                          K.U.Leuven.
\institute{Katholieke Universiteit Leuven\\
Electrical Engineering Dept. -- ACCA\\
Kard. Mercierlaan 94 -- B-3001 Heverlee -- Belgium}
\maketitle
%
% Abstract
\begin{abstract}
We present a set of C functions implementing
a distributed software voting mechanism for EPX or similar
message passing environments, and we place it within the EFTOS
framework (Embedded \FT{} Supercomputing, ESPRIT-IV Project 21012) 
of software tools for enhancing the dependability of a user application. 
The described mechanism can be used for instance
to implement restoring organs i.e., $N$\kern-1pt-modular 
redundancy systems with $N$\kern-1pt-replicated voters. 
We show that, besides structural design goals like fault transparency, 
this tool achieves replication transparency, a high degree of flexibility 
and ease-of-use, and good performance. 

\vskip4pt

\noindent
%
% Keywords
\end{abstract}
%%%%%%%%%%%%%%%%%%%%%%%%%%%%%%%%%%%%%%%%%%%%%%%%%%%%%%%%%%%%%%%%%%%%%%%%%
%%%%%%%%%%%%%%%%%%%%%% I N T R O D U C T I O N %%%%%%%%%%%%%%%%%%%%%%%%%%
%%%%%%%%%%%%%%%%%%%%%%%%%%%%%%%%%%%%%%%%%%%%%%%%%%%%%%%%%%%%%%%%%%%%%%%%%
%%
\section{Introduction}
\begin{figure}%[t]
\centerline{\psfig{figure=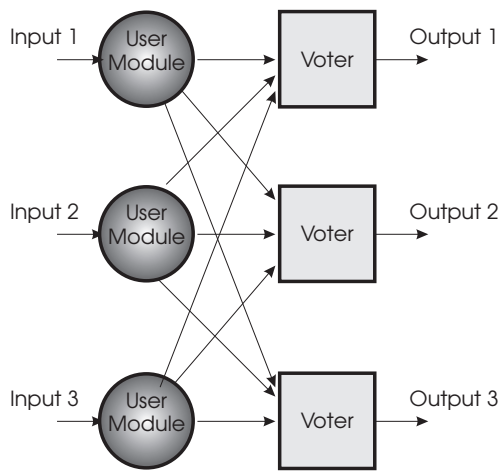,width=6.5cm}}
\caption{A ``restoring organ''~\cite{John89a} i.e., a $N$\kern-1pt-modular redundant
system with $N$ voters, when $N=3$.}
\label{ro}
\end{figure}

A well-known approach to achieve fault masking and therefore
to hide the occurrence of faults is the $N$\kern-1pt-modular redundancy
(\nmr) technique (see for instance~\cite{John89a}), valid both
on hardware and at software level.
To overcome the shortcoming of having one voter, whose failure
brings to the failure of the whole system even when each and every
other module is still running correctly, it is possible to use $N$
replicas of the voter and to provide $N$ copies of the inputs
to each replica, as described in~Fig.\ref{ro}.
This approach exhibits among others the following properties:
\begin{enumerate}
\item Depending on the voting technique adopted in the voter,
the occurrence of a limited number of faults in the inputs to the
voters may be masked to the subsequent modules~\cite{LoCE89};
for instance, by using majority voting, up to $\llcorner(N-1)/2\lrcorner$ faults can
be made transparent.
\item If we consider a pipeline of such systems, then a failing voter
in one stage of the pipeline can be simply regarded as a corrupted
input for the next stage, where it will be restored.
\end{enumerate}
The resulting system is easily recognizable to be more robust
than plain \nmr, for it does no more exhibit single points of failures.
Dependability analysis confirms intuition.
Property 2. in particular explains why such systems are
also known as ``restoring organs''~\cite{John89a}.

From the point of view of software engineering, this system 
though has two major drawbacks:
\begin{itemize}
\item Each module in the \nmr{} must be aware of and responsible for 
interacting with the whole set of voters;
\item The complexity of these interactions, which is a function increasing
quadratically with $N$, the cardinality of the voting farm, 
burdens each module in the \nmr.
\end{itemize}

As a consequence, it firstly appeared difficult to us to design a software
mechanism which, besides reaching design goals like fault transparency
(i.e., fault masking) and efficiency, were also able to achieve 
replication transparency, ease of use, and flexibility. 

\begin{figure}
\centerline{\psfig{figure=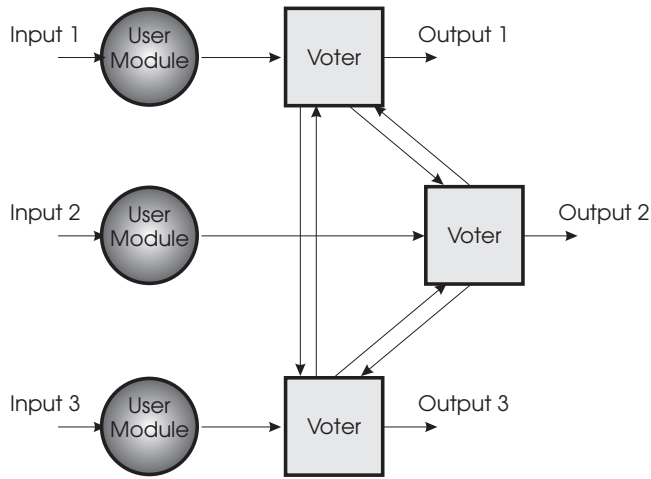,width=8.5cm}}
\caption{Structure of the EFTOS voting farm mechanism for a \nmr{} system with
$N=3$ (also known as TMR, or triple modular redundancy system.)}\label{ronew}
\end{figure}

In order to reach the full set of these requirements, we slightly
modified the design of the system as described in Fig.\ref{ronew}:
Now each module only has to interact with, and be aware of {\em one\/} voter,
regardless the value of $N$\hskip-1pt. Moreover, the complexity of such a task
is fully shifted to the voter.

We adopted this approach during the design and development of the
voting farm mechanism, a class of C functions which is part of the
EFTOS framework (Embedded \FT{} Supercomputing, ESPRIT-IV Project 21012).
In this paper we briefly draw a picture of EFTOS and place the voting farm
into it; then we describe the design of the voting farm and show how such tool
proved to fulfill the whole of our design goals and requirements. We
also describe how the user can exploit it to easily set up systems consisting
of redundant modules and based on voters.
A few notes on future developments and portability conclude
this work.

%%%%%%%%%%%%%%%%%%%%%%%%%%%%%%%%%%%%%%%%%%%%%%%%%%%%%%%%%%%%%%%%%%%%%%%%%%
\section{EFTOS and its Framework}\label{ef}
%%%%%%%%%%%%%%%%%%%%%%%%%%%%%%%%%%%%%%%%%%%%%%%%%%%%%%%%%%%%%%%%%%%%%%%%%%
The overall object of the ESPRIT-IV Project 21012 EFTOS~\cite{EFTo97a,DDLV97} 
(Embedded \FT{} Supercomputing) is to set up a software framework 
for integrating fault tolerance into embedded distributed high-performance
applications in a flexible and easy way. 
The EFTOS framework currently runs on a Parsytec CC system~\cite{Pars96a}, 
a distributed-memory MIMD supercomputer consisting of powerful processing 
nodes based on PowerPC 604 at 133MHz, dedicated high-speed links, I/O modules,
and routers. As part of the Project, this framework is currently being ported to
Microsoft Windows NT / Intel PentiumPro and
TEX / DEC Alpha platforms so to fulfill the requirements
of the EFTOS application partners.
We herein constantly refer to the version running on the CC system. 

The main characteristics of the CC system are the adoption of the 
thread processing model and of the message passing communication model: 
communicating threads exchange messages through a proprietary 
message passing library, called EPX~\cite{Pars96b} (for {\em Embedded Parallel 
extensions to uniX\/}).
A noteworthy feature of the EPX environment, which revealed to be
very useful for the voting farm, is the EPX so-called ``initial loading mechanism,''
which spawns the same executable image of the user application on each 
processing node of the user partition (a sort of ``parallel {\sf fork()}''~\cite{HaSa87}).
As a final note on EPX, we remark that it adopts the concept of 
``virtual links'' to build point-to-point connections between arbitrary threads 
within the processor pool, and that of ``local links'' to create similar
connections between threads running on a same node---the only noticeable
difference being of course in terms of performance.
Once a connection among any two threads has been set up, the involved
threads refer to it by means of a link, and use it to send and receive
messages along the same connection. {\sf Send()}'s and {\sf Receive()}'s
are synchronous and blocking---this latter attribute being a potential source of problems
from the viewpoint of fault tolerance. {\sf Receive()}'s are ``better,''
in the sense that it is possible to specify a time-out that, once reached, unblocks
the caller regardless the operation has reached completion.
Such a functionality is missing in the {\sf Send()} function.
\typeout{but there is a ``static'' version that soon will be available on TEX/ALPHA...}

Through the adoption of the EFTOS framework, the target embedded parallel 
application is plugged into a hierarchical, layered system whose structure and 
basic components are:

\begin{itemize}
\item At the lowest level, a set of parametrisable functions managing error
detection (Dtools)  and error recovery (Rtools). A typical Dtool is a
watchdog timer thread or a trap-handling mechanism;  a Rtool is e.g., a
fast-reboot thread capable of restarting a single node or a set of nodes.
These basic components are plugged into the embedded
application to make it more dependable. EFTOS supplies a number of these
Dtools and Rtools, plus an API for incorporating user-defined EFTOS-compliant
tools; 
\item At the middle level, a distributed application called {\em DIR net\/}
(detection, isolation, and recovery network)~\cite{TrDe97a} is available to coherently
combine Dtools and Rtools, to ensure consistent strategies throughout the
whole system, and to play the role of a backbone handling information 
to and from the fault tolerance elements;
%%%%To fulfill these requirements, the DIR net
%%%%makes use of processes called Manager and Agents;
\item At the highest level, these elements are combined into dependable
mechanisms i.e., methods to guarantee fault-tolerant communication, 
the voting farm mechanism, etc.
\end{itemize}

During the lifetime of the application, this framework guards
it from a series of possible deviations from the expected
activity; this is done either by executing detection, isolation, and 
reconfiguration tasks, or by means of fault masking---this latter
being provided by the EFTOS voting farm, which we are going to describe 
in the Section to follow. As a last remark, the EFTOS framework appears
to the user as a library of functions written in the C programming language.

%%%%%%%%%%%%%%%%%%%%%%%%%%%%%%%%%%%%%%%%%%%%%%%%%%%%%%%%%%%%%%%%%%%%%%%%%%
\section{The EFTOS Voting Farm}
\begin{figure}
\centerline{\psfig{figure=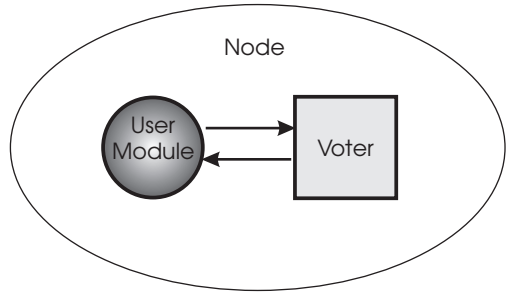,width=2.6in}}
\caption{A user module and its voter. The latter is the
only member of the farm of which the user module should be aware of:
messages will flow only between these two ends. This has been designed so
to minimize the burden of the user module and to keep it free to
continue undisturbed as much as possible.}\label{voter}
\end{figure}

%%%%%%%%%%%%%%%%%%%%%%%%%%%%%%%%%%%%%%%%%%%%%%%%%%%%%%%%%%%%%%%%%%%%%%%%%%
The basic component of our tool is the {\em voter\/} (see Fig.\ref{voter})
which we define as follows:
\begin{quote}
A voter is a local software module connected to {\em one\/}
user module and to a farm of fully interconnected fellows.
Attribute ``local'' means that both user module and voter
run on the same processing node.
\end{quote}

As a consequence of the above definition, the user module has
no other interlocutor than its voter, whose tasks are completely
transparent to the user module. It is therefore possible to model
the whole system as a simple client-server application~\cite{comer3}:
on each user module the same client protocol applies (see \S\ref{cs})
while the same server protocol is executed on every instance of the voter
(see \S\ref{ss}).

\subsection{The Client-Side of the Voting Farm}\label{cs}
Table~\ref{example} gives an example of the client-side protocol
to be executed on each processing node of the system in which a
user module runs: a well-defined, ordered list of actions has to take place
so that the voting farm be coherently declared and defined (in the sense
specified in~\cite{Stro95}),
described, activated, controlled, and queried.
In particular, {\em describing\/} a farm stands for
creating a static map of the allocation of its components; 
{\em activating\/} a farm substantially
means spawning the local voter (\S\ref{ss} will shed more light
on this); {\em controlling\/} a farm means requesting its service
by means of control and data messages;
finally, a voting farm can also be {\em queried\/} about its state, 
the current voted value, etc.

As already mentioned, the above steps have to be carried out in the 
same way on each user module: this coherency is transparently supported
by the ``initial load mechanism'' of EPX~\cite{Pars96a}.

This protocol is available to the user as a class-like collection of
functions dealing with opaque objects referenced through pointers.
A tight resemblance with the {\sf FILE} set of functions of
the standard C language library~\cite{KeRi2} has been sought so to shorten
as much as possible the user's learning time. The {\sf FILE} paradigm shows also
that, though C is certainly not the best language for object-oriented 
programming,
its support for data and function hiding\footnote{In our opinion these concepts,
available long before the conception of the \CPLUSPLUS{} programming
language~\cite{Stro95}, must have been a powerful conceptual inspirer for
this latter.}, coupled with good software practice 
can combine effectiveness, efficiency, and the elegance
of object-orientation.

The EFTOS voting farm adopts these principles and its
API and usage closely resemble those of {\sf FILE}.
It also benefits from the use of the CWEB system
of structured documentation~\cite{Knuth92} which we found an extremely
useful design tool~\cite{DeFl97b}.
\begin{table}
\begin{small}
\begin{sf}
\hrulefill
\vspace*{-12pt}
\begin{tabbing}
{\bf 001} \= VF\_control(vf, \=VF\_input(obj, sizeof(obj\_t)), \= 100000000000000000000 \= 10000000000000000000000000000 \kill\\
{\bf 1} \> VotingFarm\_t *vf; \>\>\> /* declaration */\\
{\bf 2} \> vf $\leftarrow$ VF\_open(objcmp); \>\>\> /* definition */\\
{\bf 3} \> $\forall i\in\{1,\dots,N\}$ : VF\_add(vf, node${}_i$); \>\>\> /* description */\\
{\bf 4} \> VF\_run(vf); \>\>\> /* activation */\\
{\bf 5} \> VF\_control(vf, VF\_input(obj, sizeof(obj\_t)), \\
	\>\>		   VF\_output(link),\\
	\>\>		   VF\_algorithm(VFA\_WEIGHTED\_AVERAGE),\\
	\>\>		   VF\_scaling\_factor(1.0) ); \>\> /* control */\\
{\bf 6} \> do \{\} while (VF\_error==NONE $\wedge$ VF\_get(vf)==VF\_REFUSED); \>\>\> /* query */\\
{\bf 7} \> VF\_close(vf);\>\>\> /* deactivation */
\end{tabbing}
\vspace*{-5pt}
\hrulefill
\end{sf}
\end{small}
\caption{An example of usage of the voting farm: note the resemblance
with the {\sf FILE} standard set of C language functions. {\sf objcmp()}
is a user-supplied function for comparing any two {\sf obj\_t} objects---its
role is explained in \S\ref{ss}. Note also how four messages are sent to the
local voter in Step {\bf 5}: the input to be voted, the virtual link
representing the thread to whom the voted output has to be sent, the
voting algorithm, and an optional argument pertaining the algorithm.
As a final remark, Step {\bf 6} is needed because one can only terminate a
voting farm when the broadcast of the input value is over; any attempt to
do that sooner results in a {\small\sf VF\_REFUSED} message. The loop
also checks whether a time-out has occurred during a {\small\sf VF\_get()},
in which case the global variable {\small\sf VF\_error} is set to a
value different from {\small\sf NONE}.}
\label{example}
\end{table}

\subsection{The Server-Side of the Voting Farm: the Voter}\label{ss}
The local voter thread represents the server-side of the voting farm.
After the set up of the static description of the farm (Table~\ref{example},
Step 3) in the form of an ordered list of processing node identifiers
(integer numbers greater than 0), the server-side of our application
is launched by the user by means of the {\sf VF\_run()} function. 
This turns the static representation of a farm 
into an ``alive'' (running, according to~\cite{CaGe89a}) object, the voter thread.

\begin{figure}
\centerline{\psfig{file=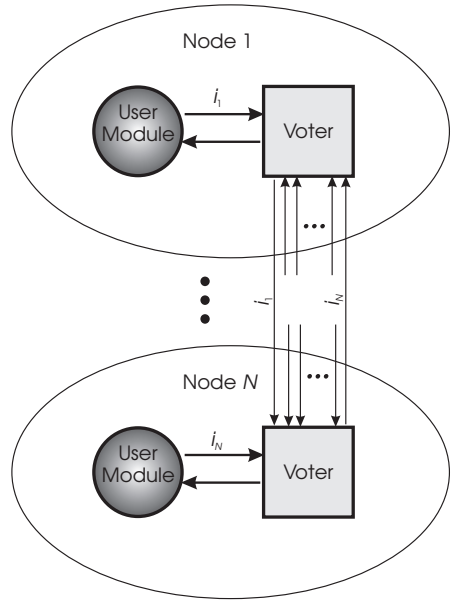,width=2.3in}}
\caption{The ``local'' input value has to be broadcasted to $N-1$ fellows,
and $N-1$ ``remote'' input values have to be collected from each of the
fellows. The voting algorithm takes place as soon as a full suite of values is
available. Note that a distributed algorithm is needed to regulate 
at all times who has the right to broadcast and who has 
to receive.}\label{vf}
\end{figure}

This latter connects to its user module via a local link and to the rest 
of the farm via virtual links. From then on, in absence of faults,
it reacts to the arrival of 
the user messages as a finite-state automaton: in particular, the input 
messages arrival triggers a number of broadcasts among the voters---as shown
in Fig.\ref{vf}---which are managed through the 
distributed algorithm described in Table~\ref{broadcast}.
When faults occur and affect up to $M<N$ voters,
no arrival for more than $\Delta t$ time units is interpreted
as the symptom of a fault. As a consequence, 
variable {\sf input\_messages} is incremented as if a message
had arrived, and its faulty state is recorded. This way we can tolerate
up to $M<N$ errors at the cost of $M\Delta t$ time units.
Note that even though this algorithm tolerates up to $N-1$ faults,
the voting algorithm may be able to cope with much less than that:
for instance, majority voting fails in the presence of 
$\llcorner(N-1)/2\lrcorner+1$ or more
faults. As another example, algorithms computing a weighted average 
of the input values consider all items
whose ``faulty bit'' is set as zero-weight values, automatically
discarding them from the average. This of course may also lead to
imprecise results as the number of faults gets larger.
\begin{table}[t]
\begin{small}
\begin{sf}
\hrulefill
\vspace*{-12pt}
\begin{tabbing}
{\bf 001} \= 1000 \= 10000000 \= 1000000000 \= 1000000000 \= 1000000 \= 1000000\kill\\
{\bf 1} \> voter\_id $\leftarrow$ who-am-i();\>\>\> /* identify yourself (voter\_id $\in\{1,\dots,N\}$) */ \\
{\bf 2} \> $\forall i$ : valid${}_i \leftarrow$ TRUE; \>\>\> /* all messages are supposed to be valid */\\
{\bf 3} \> $i\leftarrow$ input\_messages $\leftarrow$ 0;   \>\>\> /* keep track of the number of received input messages */\\
{\bf 4} \> do \{ \\
{\bf 5} \> \> Wait\_Msg\_With\_Timeout($\Delta t$);\>\>\> /* wait for an incoming message or a timeout */\\
{\bf 6} \> \> if ( Sender() == USER ) $u \leftarrow i$; \>\>\>/* $u$ points to the user module's input */\\
{\bf 7} \> \> if ( $\neg$ Timeout ) msg${}_i \leftarrow$ Receive(); \>\>\>/* read it */\\
{\bf 8} \> \> else valid${}_i \leftarrow$  FALSE; \>\>\>/* or invalidate its entry */\\
{\bf 9} \> \> $i\leftarrow$ input\_messages $\leftarrow$ input\_messages + 1;\>\>\>\>/* count it */\\
{\bf 10} \> \> if (voter\_id == input\_messages) Broadcast(msg${}_u$);\\
{\bf 11}\> \} while (input\_messages $\neq$ N);
\end{tabbing}
\vspace*{-5pt}
\hrulefill
\end{sf}
\end{small}
\caption{The distributed algorithm needed to regulate the right to
broadcast among the $N$ voters. Each voter waits for a message
for a time which is at most $\Delta t$, then it assumes a fault affected
either a user module or its voter. Function {\small\sf Broadcast()} 
sends its argument to all voters whose id is different from
{\small\sf voter\_id}. It is managed via a special sending thread
to avoid the deadlock-prone {\tt Send()}.}
\label{broadcast}
\end{table}

Besides the input value, which represents a request for voting,
the user module may send its voter a number of other requests---some 
of these are used in Table~\ref{example}, Step 5. In particular, the user
can choose to adopt a voting algorithm out of the following ones:
\begin{itemize}
\item Formalized majority voter technique,
\item Generalized median voter technique,
\item Formalized plurality voter technique,
\item Weighted averaging technique,
\end{itemize}
namely the voting techniques that were generalized in~\cite{LoCE89}
to ``arbitrary $N$\kern-1pt-version systems with arbitrary output types 
using a metric space framework.''
To use these algorithms, a metric function can be supplied by the
user when he/she ``opens'' the farm (Table~\ref{example}, Step 2):
this is exactly the same approach used in opaque C functions like
e.g., {\sf bsearch()} or {\sf qsort()}~\cite{KeRi2}.
A default metric function is also available.

Other requests include the setting of some system parameters
and the removal of the voting farm (function {\sf VF\_close()}).

%%%%%%A very special request comes with function {\sf VF\_get()}, and it is
%%%%%%a request for information on the state of a voting process
%%%%%%If the user executes this function before it gets the {\sf VF\_DONE}
%%%%%%event

The voters' replies to the incoming requests are straightforward. In particular,
a {\small {\sf VF\_DONE}} message is sent to the user module when a broadcast
has been performed; for the sake of avoiding deadlocks, one can only
control or close a farm after the {\small {\sf VF\_DONE}} message has been sent.
Any failed attempt causes the voter to send a {\small {\sf VF\_REFUSED}}
message.
This is the rationale of Step 6 in Table~\ref{example}.

Note how a function like {\sf VF\_get()} simply sets the caller in
a waiting state from which it exits either on a message arrival or
on the expiration of a time-out. Doing the other way around would have
been more error prone because of the lack-of-timeout
problem reported in Section~\ref{ef}.

%%%%%%%%%%%%%%%%%%%%%%%%%%%%%%%%%%%%%%%%%%%%%%%%%%%%%%%%%%%%%%%%%%%%%%%%%%
\section{Time and Resources Overheads of the Voting Farm.}
%%%%%%%%%%%%%%%%%%%%%%%%%%%%%%%%%%%%%%%%%%%%%%%%%%%%%%%%%%%%%%%%%%%%%%%%%%
All measurements have been performed running a restoring organ consisting
of $N$ processing nodes, $N=1,\dots,4$~\cite{EFTo97a}.
The executable file has been obtained with the {\tt ancc} C compiler using the
{\tt -O} optimization flag. %All statements that print to the screen have been
%disabled in the counted section.
During the trials the CC system was fully dedicated to the execution
of that application. 

The application has been executed in four runs, each of which
has been repeated fifty times, increasing the number of voters from
1 to 4.  Wall-clock times have been collected.
Averages and standard deviations are shown in Table~\ref{run1}.

\begin{table}
\centerline{\begin{tabular}{|r|c|c|}
\hline
number of nodes	& \ \ average\ \ \ & standard deviation \\ \hline
\hbox to 25pt{1}	& 0.000615	& 0.000006 \\ 
\hbox to 25pt{2}	& 0.001684	& 0.000022 \\ 
\hbox to 25pt{3}	& 0.002224	& 0.000035 \\ 
\hbox to 25pt{4}	& 0.003502	& 0.000144 \\ 
\hline
\end{tabular}
}
\caption{Time overhead of the voting farm for one to four node systems. 
The unit is seconds.}
\label{run1}
\end{table}

As of the overhead in resources, $N$ threads have to be spawned,
and $N$ local links are needed for the communication between each
user module and its local voter.
The network of voters calls for another 
$\frac{N\times \left( N-1 \right) }{2}$ virtual links.

%%%%%%%%%%%%%%%%%%%%%%%%%%%%%%%%%%%%%%%%%%%%%%%%%%%%%%%%%%%%%%%%%%%%%%%%%%
\section{Conclusions}
%%%%%%%%%%%%%%%%%%%%%%%%%%%%%%%%%%%%%%%%%%%%%%%%%%%%%%%%%%%%%%%%%%%%%%%%%%
A flexible, easy to use, efficient mechanism for software voting 
in message passing systems has been described. 
The tool, currently running on a Parsytec CC system,
has been designed with portability in mind and is actually being
ported to a PentiumPro/Windows NT and a Alpha/TEX platform. A special,
``static'' version is being developed for this latter, which adopts
the mailbox paradigm as opposed to message passing via virtual links.

We are currently considering some additional improvements and extensions of
the voting farm, including the possibility for the user to supply
voting algorithms of his/her choice, and a tighter link with the
EFTOS fault tolerance backbone: in particular, the voting farm
will inform the DIR net about the state of the voting sessions, namely
who failed, and on which nodes this happened. This information shall be
exploited by the ``error diagnosis engine'' \cite{TrDe97a} of the DIR net.
\typeout{Non piu' "un voter per nodo"}
\typeout{Man pages sono disponibili..}

%%%%%%%%%%%%%%%%%%%%%%%%%%%%%%%%%%%%%%%%%%%%%%%%%%%%%%%%%%%%%%%%%%%%%%%%%%%
\paragraph{Acknowledgments.}
This project is partly sponsored by 
     an FWO Krediet aan Navorsers, 
     by the Esprit-IV Project 21012 EFTOS, 
 and by COF/96/11. 
Vincenzo De Florio is on leave from Tecnopolis CSATA Novus Ortus.  
Geert Deconinck has a grant from the Flemish Institute for
the Promotion of Scientific and Technological Research in Industry (IWT).
Rudy Lauwereins is a Senior Research Associate of the Fund for Scientific 
Research - Flanders (Belgium).

%
% ---- Bibliography ----
%

\end{document}